\documentclass[10pt,twocolumn,pra,amssymb,amsmath,nobibnotes,aps,showpacs]{revtex4-1}
\pdfoutput=0
\usepackage{booktabs}
\usepackage{graphics} \usepackage{graphicx} \usepackage{braket} \usepackage{physics} %\usepackage{float}  % Advanced table configurations
\usepackage[hyperindex,breaklinks]{hyperref}

\usepackage{mathrsfs}
\usepackage{tabularx}

\usepackage{multirow}

\usepackage{rotating}
\usepackage{epstopdf}

\begin{document}

\title{Hyperfine Structure of the $B^3\Pi_1$ State and Predictions of Optical Cycling Behavior in the $X\rightarrow B$ transition of TlF}

%\author{E. B. Norrgard$^{1}$, N. Sitaraman$^{1}$, and D. DeMille$^{1}$}

\author{E. B. Norrgard}

\email[Electronic address: ]{eric.norrgard@yale.edu}

\author{E. R. Edwards}

\author{D. J. McCarron}

\author{M. H. Steinecker}

\author{D. DeMille}
\email{david.demille@yale.edu}

%\altaffiliation[Current address: ]{Department of Physics, Cornell University, 142 Sciences Dr., Ithaca, New York 14853, USA}

\affiliation{Department of Physics, Yale University, P.O. Box 208120, New Haven, Connecticut 06520, USA}
\author{Shah Saad Alam}

\author{S. K. Peck}

\author{N. S. Wadia}

\author{L. R. Hunter}
\email{lrhunter@amherst.edu}

\affiliation{Physics Department, Amherst College, Amherst, Massachusetts 01002, USA}

\begin{abstract}
  The rotational and hyperfine spectrum of the $X^1\Sigma^+ \rightarrow B^3\Pi_1$ transition in TlF molecules was measured using laser-induced fluorescence from both a thermal and a cryogenic molecular beam.  Rotational and hyperfine constants for the $B$ state are obtained.  The large magnetic hyperfine interaction of the Tl nuclear spin leads to significant mixing of the lowest $B$ state rotational levels.  Updated, more precise measurements of the $B\rightarrow X$ vibrational branching fractions are also presented.  The combined rovibrational branching fractions allow for the prediction of the number of photons that can be scattered in a given TlF optical cycling scheme.
\end{abstract}

\maketitle
%\pacs{<33.15Pw, 33.20Lg, 37.10Mn>}

 %Introduction
 \section{Introduction}
The $X^1\Sigma^+$ state of thallium monofluoride (TlF) has long been a platform for precision measurements of parity- and time-reversal symmetry violation \cite{Hinds1980,Wilkening1984,Cho1991}, with high potential for discovery of new physics \cite{Khriplovich1997}.  In particular, the high mass of Tl coupled with the high polarizability of the molecule make this system ideal for measuring the Schiff moment of the Tl nucleus  \cite{Schiff1963}.

 Optical cycling \cite{Shuman2009} is a potentially powerful tool for enhancing the signal in a symmetry violation measurement.  Unit-efficiency detection of the internal state is possible when the number of photons scattered per molecule exceeds the reciprocal of the total light collection efficiency (geometric and detector efficiencies).   Optical cycling also allows for the application of large optical forces, which can be useful for precision measurements. Transverse cooling can decrease beam divergence (and increase downstream flux) \cite{Shuman2010}, and longitudinal slowing and  trapping can increase effective interaction times \cite{Barry2014}, leading to improved energy resolution.

Optical cycling requires the ability to optically couple a subspace of ground and excited states, such that the excited states rarely decay to uncoupled ground states. Recently, we proposed the TlF $X^1\Sigma^+ (v_g\!=\!0)$\,$\rightarrow$\,$B^3\Pi_1(v_e\!=\!0)$ transition (where $v_g$ ($v_e$) is the ground (excited) state vibrational quantum number) as a candidate for optical cycling and laser cooling \cite{Hunter2012}, as this transition has a sufficiently short excited state lifetime ($\tau$\,$=$\,$99$\,ns) and favorable vibrational branching fractions to form a quasi-closed optical cycle.  In TlF, the $B^3\Pi_1$ state was expected to have resolved -- and potentially very large -- hyperfine (HF) structure \cite{Hunter2012}.  The HF interaction in the excited state can lead to mixing of rotational states with different quantum numbers $J$; this in turn can break the usual rotational selection rules, leading to branching to additional ground rotational levels which must be coupled to the optical cycle.  Hence, it is crucial to characterize the rotational and HF structure of the excited state to understand and control rotational branching.

High-resolution microwave spectroscopy \cite{Hoeft1970} has provided a detailed and precise understanding of the $X$ state HF and rotational energies.   Low-resolution spectroscopy with a pulsed UV laser by Wolf and Tiemann \cite{Wolf1987} allowed for determination of  rovibrational energies of the $B^3\Pi_1$ state.  In this paper, we present high-resolution laser spectroscopy of the $X^1\Sigma^+(v_g=0)$\,$\rightarrow$\,$B^3\Pi_1(v_e\!=\!0)$ transition. We clearly identify rotational lines associated with $B^3\Pi_1(v_e=0)$ states from $J= 1$ -- 70.  The $B$ state HF structure is fully resolved for $J \lesssim 51$ and is fit to a standard Hamiltonian  to determine the parameters describing the HF interaction. These data allow a full characterization of the HF structure in the $B$ state, including effects of HF-induced rotational state mixing.  In addition, we present improved vibrational branching measurements from $v_e=0$.  Together, these data are used to quantitatively predict the number of photons that may be scattered on the $X^1\Sigma^+$\,$\rightarrow$\,$B^3\Pi_1$ transition of TlF for various cycling schemes.

%Experiment
\section{Experimental Details}
To allow access to and identification of a large range of rotational states, we make observations using both a thermal beam source and a cryogenic buffer gas beam source.
The thermal oven source is the same as in Ref.\,\citep{Hunter2012}. It is described in brief here. The measurements are made in a stainless steel molecular beam apparatus maintained at a pressure of about $10^{-6}$\,Torr. The beam itself is created by heating a stainless steel oven containing TlF to temperatures of 688-733 K. The beam emerges from the oven through four ceramic tubes. The tubes precollimate the beam, which is then further collimated by an aperture located about 30 cm from the oven.

The cryogenic buffer gas beam source is nearly identical to that of \cite{Barry2011}.  A solid target is made by melting TlF powder in a copper crucible under vacuum.  The crucible is mounted inside a copper cell and  held at roughly 4\,K by a pulse tube refrigerator.  TlF molecules are produced by laser ablation of the target with 10\,ns, 25\,mJ pulses of 1064\,nm light, and are extracted from the cell by a flow of typically 5\,sccm of cryogenic helium buffer gas.  The molecular beam then propagates through a region held at roughly $10^{-7}$\,Torr.

The $X(0)$\,$\rightarrow$\,$B(0)$ transition occurs at wavelength 271.7\,nm.
A CW, single-frequency, tunable 1087\,nm fiber laser is frequency doubled twice to produce roughly 20\,mW of CW 271.7\,nm light, using commercial resonant bow-tie cavities.  The fiber laser frequency is locked by monitoring its transmission through a scanning Fabry-Perot cavity; the Fabry-Perot cavity length is in turn stabilized by simultaneously monitoring the transmission of a frequency-stabilized helium-neon laser.

In both molecular beam setups, the 271.7\,nm laser light is directed to intersect the molecular beam perpendicular to the direction of molecule motion.  Laser-induced fluorescence is collected onto a photomultiplier tube (PMT) in photon counting mode.   In the thermal source, the fluorescence is collimated, passed through an interference filter, and then spatially filtered and collected.  In the cryogenic source, the fluorescence is transferred to the PMT by a light pipe, followed by an interference filter.

%HF Structure Measurement

%Modeling HF Structure, State Content

%Franck-Condon Factors

%Predicted Cycling Behavior

%X state HF Remixing

%Conclusion

\section{Hyperfine and Rotation Hamiltonian}\label{sec:hyperinehamiltonian}
\subsection{Quantum Numbers in the $X-B$ Spectrum}

Thallium has two common isotopes, $^{203}$Tl (30\,\% natural abundance) and $^{205}$Tl (70\,\%), both with nuclear spin $I_1$\,$=$\,$1/2$ \cite{Lide1997}.  Fluorine's only isotope is $^{19}$F, also with nuclear spin $I_2$\,$=$\,$1/2$. We describe the $B$ state of TlF using the Hund's case (c) basis and the  coupling scheme:
\begin{equation}\label{eq:coupling1}
\begin{aligned}
  \boldsymbol{F_1}&=&\boldsymbol{J}+\boldsymbol{I_1},\\
   \boldsymbol{F}&=&\boldsymbol{F_1}+\boldsymbol{I_2},
   \end{aligned}
\end{equation}
where $J$ is the total angular momentum of the molecule less nuclear spin.  Hence each rotational state with quantum number $J$ has associated HF states with $F_1$\,$=$\,$J \pm \frac{1}{2}$ and  $F$\,$=$\,$J-1$, $J$, $J$, and $J+1$.

The Hund's case (c) basis kets $\ket{c}$ are:
\begin{equation}\label{eq:vectors}
  \ket{c}=\ket{J,\Omega,I_1,F_1,I_2,F,m_F,P}.
     \end{equation}
Here, $\Omega$ is the projection of $J$ on the internuclear axis, $m_F$ is the projection of $F$ in the lab frame, and $P= \pm 1$ is the state parity.  Following the convention of Herzberg \cite{Herzberg1950}, we refer to states with $P$\,$=$\,$(-1)^J$ as $e$-parity and $P = (-1)^{J+1}$ as $f$-parity states.  As described below, the large HF interactions in the $B$ state mix neighboring rotational levels.  We use rotational quantum number $J$ to describe states in the basis of Eq.\,\ref{eq:vectors}, and label energy eigenstates in the case of large mixing by $\tilde{J}$ (i.e.\ $\tilde{J}=J$ in the absence of HF mixing).  We denote the rotational quantum number in the ground state by $J_g$.

\subsection{Magnetic Hyperfine}

The largest HF effect is expected to be due to the magnetic HF interaction, described by the Hamiltonian  $H_{\rm{mhf}}$:
\begin{equation}\label{eq:mhf1}
  H_{\rm{mhf}} = a\boldsymbol{I}\cdot\boldsymbol{L}+b\boldsymbol{I}\cdot\boldsymbol{S}+cI_z S_z,
\end{equation}
where $I$\,$=$\,$I_1$ or $I_2$; $L$ is the total electron orbital angular momentum; and $S$ is the total electron spin. The lower-case subscript corresponds to coordinates in the molecule fixed-frame, with $\hat{z}$ along the internuclear axis. In the limit of negligible coupling to other electronic states via  $H_{\rm{mhf}}$, we may write the effective Hamiltonian in the form
\begin{equation}\label{eq:mhf2}
\begin{aligned}
  H_{\rm{mhf}}^{\rm{eff}} =& \bigl( aL_z+(b+c)S_z \bigr) I_z,\\
  =& h_\Omega I_z.
  \end{aligned}
\end{equation}
Here, $h_\Omega$\,$=$\,$a \langle L_z\rangle + (b+c)\langle S_z\rangle$, where $\langle\rangle$ corresponds to the expectation value of the operators in the electronic state of interest.
% Note that for $^3\Pi_1$ and $^1\Pi_1$ states, $\expval{\boldsymbol{S_z}} = \Sigma = 0$, $\langle \boldsymbol{L_z}\rangle =\Lambda =1$ and $h_1 \equiv a$.  This is a fair estimate for the so-called $B^3\Pi_1$ state of TlF, which according to \emph{ab initio} calculations has 84\,\% $^3\Pi_1$ character, 12\,\% $^1\Pi_1$ character, and 4\,\% $^3\Sigma^+_1$ character \cite{Zou2008}.  Using this approximation, we find f
For $I$\,$=$\,$I_1$,

 \begin{equation}\label{eq:mhfI1}
 \begin{aligned}
  &\matrixelement{J,\Omega,F_1,F,m}{H_{\rm{mhf}}\rm{(Tl)}}{J^\prime,\Omega^\prime,F_1^\prime,F,m}  \\
&= h_1(\text{Tl}) (-1)^{J+J^\prime +F_1 +I_1-\Omega}\delta_{F_1,F_1^\prime} \\
& \quad \times[(2J+1)(2J^\prime+1)I_1(I_1+1)(2I_1+1)]^{1/2}\\
&\quad \times\begin{Bmatrix}
I_1 & J^\prime & F_1 \\
J & I_1 & 1
\end{Bmatrix} \mqty( J & 1 & J^\prime \\ -\Omega & 0 & \Omega^\prime).
\end{aligned}
\end{equation}

Similarly for $I= I_2$,

\begin{equation}\label{eq:mhfI2}
 \begin{aligned}
  &\matrixelement{J,\Omega,F_1,F,m}{H_{\rm{mhf}}\rm{(F)}}{J^\prime,\Omega^\prime,F_1^\prime,F,m}  \\
 &= h_1(\text{F})(-1)^{2F_1^\prime+F+2J-\Omega+1+I_1+I_2} \\
 &\quad \times \begin{Bmatrix}
I_2 & F_1^\prime & F \\
F_1 & I_2 & 1
\end{Bmatrix}\begin{Bmatrix}
J^\prime & F_1^\prime & I_1 \\
F_1 & J & 1
\end{Bmatrix} \mqty( J & 1 & J^\prime \\ -\Omega & 0 & \Omega^\prime)\\
&\quad \times \bigl[(2F_1+1)(2F_1^\prime+1)(2J+1)(2J^\prime+1)\\
&\quad \quad \times I_2(I_2+1)(2I_2+1)\bigr]^{1/2} .
  \end{aligned}
\end{equation}

\subsection{Nuclear Spin-Rotation}
The effective nuclear spin-rotation Hamiltonian is of the form $H_{\rm{nsr}}$\,$=$\,$c_I (\boldsymbol{I}\cdot\boldsymbol{J})$, where again $I =I_1$ or $I_2$.  This arises from two physical mechanisms.  The first is the coupling of the rotational motion of the nuclei to the nuclear spin magnetic moments.  For an electronic state which is not strongly perturbed by other nearby states, this contribution dominates, and $c_I$ is quite small (for example, $c_I$(Tl)$\,$=$\,$126.03\,kHz and $c_I$(F)$\,$=$\,$17.89\,kHz in the $X^1\Sigma$ state in TlF \cite{Hinds1980}).
The second contribution arises from second-order rotational coupling to other electronic states  \cite{Okabayashi2012}. %, given by
%\begin{equation}\label{eq:nsr2ndterm}
 % c_I^{(2)} = 4B \sum_{n}\frac{a_{0 n}\matrixelement{0}{L_x}{n}^2}{E_n-E_0},
%\end{equation}
%where $B$ is the rotational constant, $a_{0 n} = 2 \mu_B g_N \mu_N \matrixelement{0}{r^{-3}}{n}$ characterizes the strength of the orbital electron-nuclear spin magnetic hyperfine interaction,  and $\ket{0}$ and $\ket{n}$ are the electronic state of interest and perturbing states, respectively \cite{Okabayashi2012}.
 This contribution likely dominates in the $B$ state, where levels with $^3\Delta$ and $^3\Sigma$ character are predicted to lie near in energy %, though they have not been observed experimentally
  \cite{Balasubramanian1985}.
  %Additionally, the small (4\,\%) $^3\Sigma^+$ character of the $B$ state allows it to interact with the rest of the $^3\Pi$ manifold; these states may be only $\approx 1000$\,cm$^{-1}$ away from the $B$ state \cite{Zou2008}.
  In such a situation, $c_I$ can be significantly larger  than when the main contribution comes from the first mechanism; for example, in  $^{195}$PtF, $c_I$(Pt)$\,$=$\,$3.11\,MHz \cite{Okabayashi2012}.

The nuclear spin-rotation matrix elements can be written as follows.  For $I$\,$=$\,$I_1$:
 \begin{equation}\label{eq:nsr1}
 \begin{aligned}
  &\matrixelement{J,\Omega,F_1,F,m}{H_{\rm{nsr}}\rm{(Tl)}}{J^\prime,\Omega^\prime,F_1^\prime,F,m} \\
  & = c_I(\text{Tl})(-1)^{J+F_1+I_1}\delta_{F_1,F_1^\prime}\delta_{J,J^\prime}\begin{Bmatrix}
I_1 & J & F_1 \\
J & I_1 & 1
\end{Bmatrix}\\
& \quad \times [(J(J+1)(2J+1)I_1(I_1+1)(2I_1)+1)]^{1/2}.
\end{aligned}
\end{equation}
For $I= I_2$:
\begin{equation}\label{eq:nsr2}
 \begin{aligned}
  &\matrixelement{J,\Omega,F_1,F,m}{H_{\rm{nsr}}\rm{(F)}}{J^\prime,\Omega^\prime,F_1^\prime,F,m} \\
  &= c_I(\text{F}) (-1)^{2F_1^\prime+F+J+I_1+I_2+1}\delta_{J,J^\prime}\\
  &\quad \times \begin{Bmatrix}
I_2 & F_1^\prime & F \\
F_1 & I_2 & 1
\end{Bmatrix}\begin{Bmatrix}
J^\prime & F_1^\prime & I_1 \\
F_1 & J & 1
\end{Bmatrix}\\
&\quad\times \bigl[(2F_1+1)(2F_1^\prime+1)J(J+1)(2J+1)\\
&\quad \quad \times I_2(I_2+1)(2I_2+1)\bigr]^{1/2}.
\end{aligned}
\end{equation}
Because there are discrepancies in the literature about the explicit form of these HF matrix elements, we derive them in Appendix C.

\subsection{Additional Terms}

Equations \ref{eq:mhfI1} and \ref{eq:mhfI2} are only strictly valid for an isolated electronic level.  Perturbations by a nearby level can lead to extra terms in the effective Hamiltonian with the $J$-dependence of centrifugal terms, modeled by substituting $h_1 \rightarrow h_1 +h_{1D} J(J+1)$ \cite{Okabayashi2003}.  However, the diagonal  matrix elements of the centrifugal magnetic HF and nuclear spin-rotation Hamiltonians are identical, and the effects of off-diagonal elements are too small to be discernable in our data.  We choose to constrain $h_{1D}$\,$=$\,$0$, and treat the fit constants $c_I$(Tl) and $c_I$(F) as empirical combinations of the two effects.

Rotational energy is modeled using the standard effective Hamiltonian $H_{\rm{rot}}$ \cite{Herzberg1950}:
\begin{equation}\label{eq:Rotational Hamiltonian0}
  H_{\rm{rot}} = B_0\boldsymbol{J}^2-D_0\boldsymbol{J}^4+H_0\boldsymbol{J}^6\dots ,
\end{equation}
with diagonal matrix elements
\begin{equation}\label{eq:Rotational Hamiltonian}
 \begin{aligned}
  &\matrixelement{J,\Omega,F_1,F,m}{H_{\rm{rot}}\rm{(F)}}{J,\Omega,F_1,F,m} \\
  &\quad= B_0J(J+1)-D_0(J(J+1))^2+H_0(J(J+1))^3\dots .
  \end{aligned}
\end{equation}

  The $B$ state HF/rotation structure is expected to depend on the isotopologue and the $e$/$f$ parity. We therefore fit the parity states and each Tl isotope separately. This is equivalent to the analysis of Ref.\,\cite{Okabayashi2012}, which used the substitution $h_1$\,$\rightarrow$\,$h_1 \pm h_{1q}$ and and $c_I$\,$\rightarrow$\,$c_I \pm c_{Iq}$, with the upper (lower) sign used for the $e$- ($f$-) parity states.

\section{Observed Spectral Features}

\subsection{Line Identification}
\begin{figure*}[t]
\centering
\includegraphics[width= \linewidth]{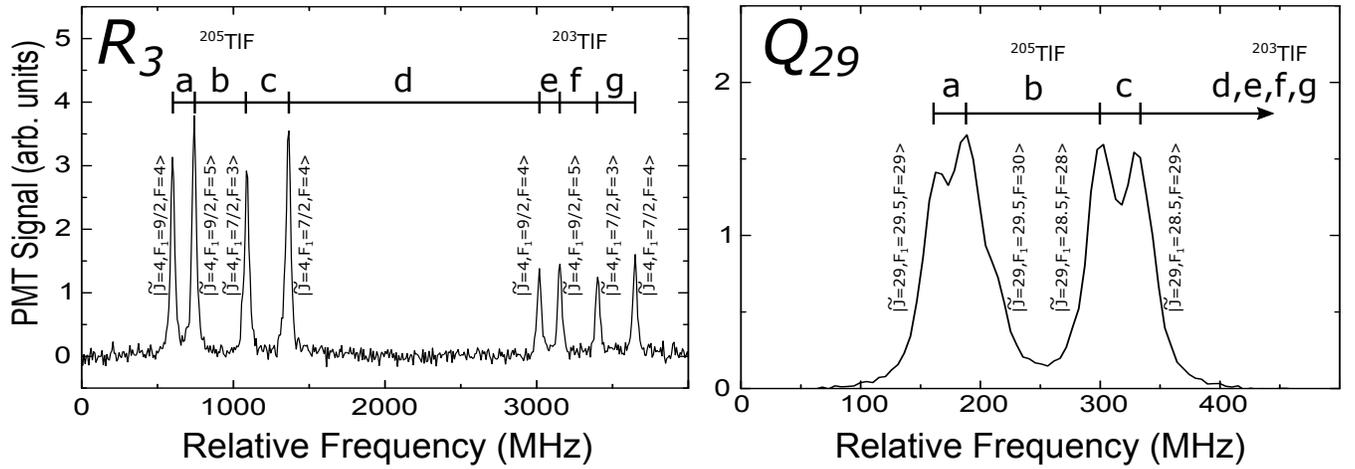}
        \caption{PMT signal vs laser frequency in the thermal source.  The scan over the $R_3$ lines (left) shows the labeling scheme for observed isotope/HF splittings used in the $P$, $Q$ and $R$ branch lines.  The basic pattern of HF structure repeats with decreasing splittings as $\Tilde{J}$ increases, as observed in the $Q_{29}$ line (right).   }\label{fig:RandQscans}
\end{figure*}

The HF structure of the ground $X^1\Sigma^+$ state is unresolved ($\sim$\,100\,kHz) in our optical spectra.  In the excited $B^3\Pi_1$ state, we expect 8 well-split isotope/HF sublevels for each rotational level $\tilde{J}$ and parity $P$, corresponding to $F_1$\,$=$\,$\tilde{J}\pm\frac{1}{2}$, $F$\,$=$\,$F_1\pm\frac{1}{2}$, and the two Tl isotopes.  We label the 7 splittings between the 8 lines as $a, b, \dots g$ as shown in Fig.\,\ref{fig:RandQscans}.    In most cases, we easily identify lines associated with the two isotopologues, since their intensities are proportional to the isotopic abundance: we associate splittings $a,b,c$ with $^{205}$Tl; $e,f,g$ with $^{203}$Tl; and $d$ with the gap between the isotopologues. For a given isotopologue, the largest splitting (associated with the Tl nuclear spin projection) is $b$ or $f$. The doublets separated by $a, c$ ($e, g$) then correspond to the $^{19}$F nuclear spin projection for $^{205}$TlF ($^{203}$TlF).  As described in Section \ref{sec:largeJ1} below, in $\Tilde{J} = 1$ only, the separation of the levels due to the Tl spin projection is larger than the isotope shift;  we use the convention of labeling the splitting between the highest $^{205}$TlF HF level and the lowest $^{203}$TlF HF level as $d$, and thus $d < 0$ for transitions to $\Tilde{J}$\,$=$\,$1$.

%\begin{table}[b]
%  \centering
%  \begin{ruledtabular}
%  \begin{tabularx}{\linewidth}{c c c }
%Parameter& Ref.\,\cite{Tiemann1988}& This Work\\ \hline
%$B$ & 6688.7(11) &6686.7(11)\\
%$D$ &0.01096(13) & 0.0110(2) \\
%$H$ &$-9.0(6)\cdot10^{-8}$ &$-6.1(1)\cdot10^{-8}$ \\

%\end{tabularx}
%  \caption{Rotational constants for $B^3\Pi_1 (v_e=0)$. Numbers in parentheses are the $1\sigma$ confidence intervals. All values are in MHz.}\label{tab:rotational constants}
%  \end{ruledtabular}
%\end{table}

\begin{table*}
  \centering
\footnotesize
\begin{ruledtabular}
  \begin{tabularx}{\textwidth}{c || c | c c c c c c c c|| c| c c c c c c c c}
$\tilde{\textbf{J}}$ & \textbf{R} & \textbf{a} &\textbf{b}&\textbf{c}&  \textbf{d} &\textbf{e}&\textbf{f}&\textbf{g}& &\textbf{P} & \textbf{a} &\textbf{b}&\textbf{c}&  \textbf{d} &\textbf{e}&\textbf{f}&\textbf{g}\\ \hline
21& 20 &37 &354 &42 &  &  & \textit{} & \textit{} & \textit{}& & \\
\dots & & & & & & & & & & &\\
16& 15 &50 &303 &60 &  &  & \textit{} & \textit{} & \textit{} & &\\
\dots & & & & & & & & & & &\\
11& 10 &66 &274 &83 &  &  & \textit{} & \textit{} & \textit{} & &\\
\dots & & & & & & & & & & &\\

9& 8 &\textit{91} &\textit{272} &\textit{99} &  &  & \textit{} & \textit{} & \textit{}& & \\
8& 7 &\textit{88} &\textit{270} &\textit{110} &\textit{2203}&\textit{80}&\textit{224} &\textit{113} &  & \textit{} & \textit{} & \textit{} & &\\
7& 6 &\textit{93} &\textit{269} &\textit{128} &\textit{2157}&\textit{93}&\textit{224}&\textit{128}&  &  & \textit{} & \textit{} & \textit{} \\

6& 5 &105 &297 &153 &2108&113&231&150&  &  & \textit{} & \textit{} & \textit{} \\
5& 4 &121 &304 &189&2050&125&247&185 &  &  & \textit{} & \textit{} & \textit{} \\
4& 3 &147 &343 &255 & 1909&147&266&249 & & 5 & \textit{142} & \textit{379} & \textit{250}& \textit{1813}& \textit{151}& \textit{254}& \textit{247}\\
3& 2 &178 &391 &341 &1735&179&302&335&  & 4 & 175 & 388 & 341 & 1646&188&309&332 \\
2& 1 &227 &463 &580& 1378&229&326&579&  & 3 & \textit{256} & \textit{475} & \textit{562}& \textit{1340}& \textit{225}& \textit{315}& \textit{560} \\
1& 0 &\textit{548} &\textit{13524} &\textit{304} & & & &  \textit{289}& & 2 & \textit{545} & \textit{13523} & \textit{305}& \textit{-11711}& \textit{582}& \textit{13431}& \textit{309}\\
\end{tabularx}

  \caption{Observed splittings of $e$-parity levels (MHz). Regular (italic) type indicates splittings measured in the thermal (cryogenic) beam.}\label{tab:eparity}
  \end{ruledtabular}
\end{table*}
\normalsize

\begin{table*}[t]
\centering
\footnotesize
\begin{ruledtabular}
\begin{tabularx}{\linewidth}{c||c| c c ccccc||c| ccc|| c |ccc}
$\tilde{\textbf{J}}$ &\textbf{Q} & \textbf{a} &\textbf{b}&\textbf{c}&  \textbf{d} &\textbf{e}&\textbf{f}&\textbf{g}& \textbf{O} & \textbf{h} & \textbf{i}& \textbf{j}&\textbf{S} & \textbf{k} & \textbf{l}& \textbf{m}\\ \hline
70&70& &   & &  & &177 & &  & & & &\\
69&69& &   & &  & &176 & &  & & & &\\
68&68& &179& &  & &171 & &  & & & &\\
67&67& &   & &  & &173 & &  & & & &\\
66&66& &175& &  & &168 & & & & & &\\
65&65& &176& &  & &168 & &  & & & &\\
64&64& &175& &  & &167 & & & & & &\\
63&63& &184& &  & &166 & &  & & & &\\
62&62& &177& &  & &165 & & & & & &\\
61&61& &170& &  & &164 & &  & & & &\\
60&60& &168& &  & & & & & & & &\\
59&59& &157& &  & & & &  & & & &\\
\dots & & & & & & & & & & & & & & &\\
51&51&17 &139& 17&  & & & & & & & & \\
\dots & & & & & & & & & & & & & & &\\
34&34&27 &111& 26&  & & & & & & & & \\
\dots & & & & & & & & & & & & & & &\\
29&29 &26 &114 &26 &  & & & & & & & & \\
\dots & & & & & & & & & & & & & & &\\
27&27 &25 &113 &32 &  & & & &  & & & &\\
\dots & & & & & & & & & & & & & & &\\
25&25 &30 &108 &29 &  & & & & & & & & \\
\dots & & & & & & & & & & & & & & &\\
23&23 &37 &110 &42 &  & & & & & & & & \\
\dots & & & & & & & & & &  & & & & &\\
20&20 &36 &110 &45 &  & & & & & & & & \\
\dots & & & & & & & & & & & & & & &\\
17&17 &46 &115 &50 &  & & & &  & & & &\\
16&16 &48 &115 &54 &  & & & &  & & & &\\
\dots & & & & & & & & & & & & & & &\\
14&14 &56 &121 &64 &  & & & &  & & & &\\
13&13 &59 &133 &70 &  & & & &  & & & &\\
12&12 &63 &131 &76 &  & & & &  & & & &\\
11&11 &68 &140 &80 &  & & & &  & & & &\\
10&10 &\textbf{74} &\textbf{157} &\textbf{78} &  & & & &  & & & &\\
9&9 &75 &155 &100 &  & & & &  & & & &\\
8&8 &\textbf{90} &\textbf{164} &111 &  & & & &  & & & &\\
7&7 &\textbf{93} &194 &\textbf{116} &  & & & &  & & & &\\
6&6 &\textbf{107} &\textbf{221} &150 &  &           & & & & & & &4&               &               & 154\\
5&5 &\textbf{129} &\textbf{255} &\textbf{179} &  & \textit{} & & & & & & &3&  \textit{180}&  \textit{2430}&  \textit{184}\\
4&4 &152 &302 &245 &  & \textit{} & & & & & & &2&  \textit{244}&  \textit{2355}&  \textit{235}\\
3& 3&\textbf{182} &372 &344 &  & \textit{} & & & & & & &1&  \textit{367}&  \textit{2229}&  \textit{347} \\
2& 2&\textbf{233} &\textbf{472} &\textbf{568} &  & \textit{} & & & & & & &0&  \textit{561}&  \textit{1948}&  \textit{578} \\
1&1 &571 &13522 &\textbf{316}& \textit{-11361}& \textit{553}& & &  3 & \textit{312} & \textit{2269} & \textit{309}& & \\
\end{tabularx}\caption{Observed splittings of $f$-parity levels (MHz). Regular (italic) type indicates splittings measured in the thermal (cryogenic) beam. Bold indicates splittings of tentative line assignments based on comparisons of overlapping lines present in both beams. }\label{tab:fparity}
\end{ruledtabular}
\end{table*}
\normalsize

The $R$ and $P$ branch transitions are spectroscopically isolated and hence relatively easy to identify for $\Tilde{J} > 1$.  For these states, 8-line multiplets similar to that displayed in Fig.\,\ref{fig:RandQscans} are found approximately centered on the locations predicted from the Dunham coefficients of Tiemann \cite{Tiemann1988}.  Our measured splittings for the various values of $e$-parity excited states are listed in Table \ref{tab:eparity}.

Assignment of the $Q$ branch transitions is more challenging.  Due to the near equality of the rotational constants of the $X$ and $B$ states, the $Q$ branch transitions are generally not clearly separated.  All of the lines between $Q_1$ and $Q_{60}$ (approximately 480 individual HF transitions) are contained in a frequency range of about 21\,GHz (see Appendix D).
We initially identified many of the $Q$ branch multiplets for $Q_{11}$--$Q_{34}$ using rotational constants $B_0$, $D_0$, and $H_0$ from Tiemann \cite{Tiemann1988}, then fine-tuned these parameters to obtain good agreement with all the identifiable $Q$ branch lines ($Q_{11}$--$Q_{68}$).  Because of their lower abundance and correspondingly smaller signals, no similar identification was possible for the majority of the $^{203}$TlF $Q$ branch lines.  Assuming the same rotational constants (scaled for reduced mass and including an isotopic offset), we have made some tentative assignments of $^{203}$TlF $Q_{60}$--$Q_{71}$.

\begin{figure*}
\centering
\includegraphics[width= \textwidth]{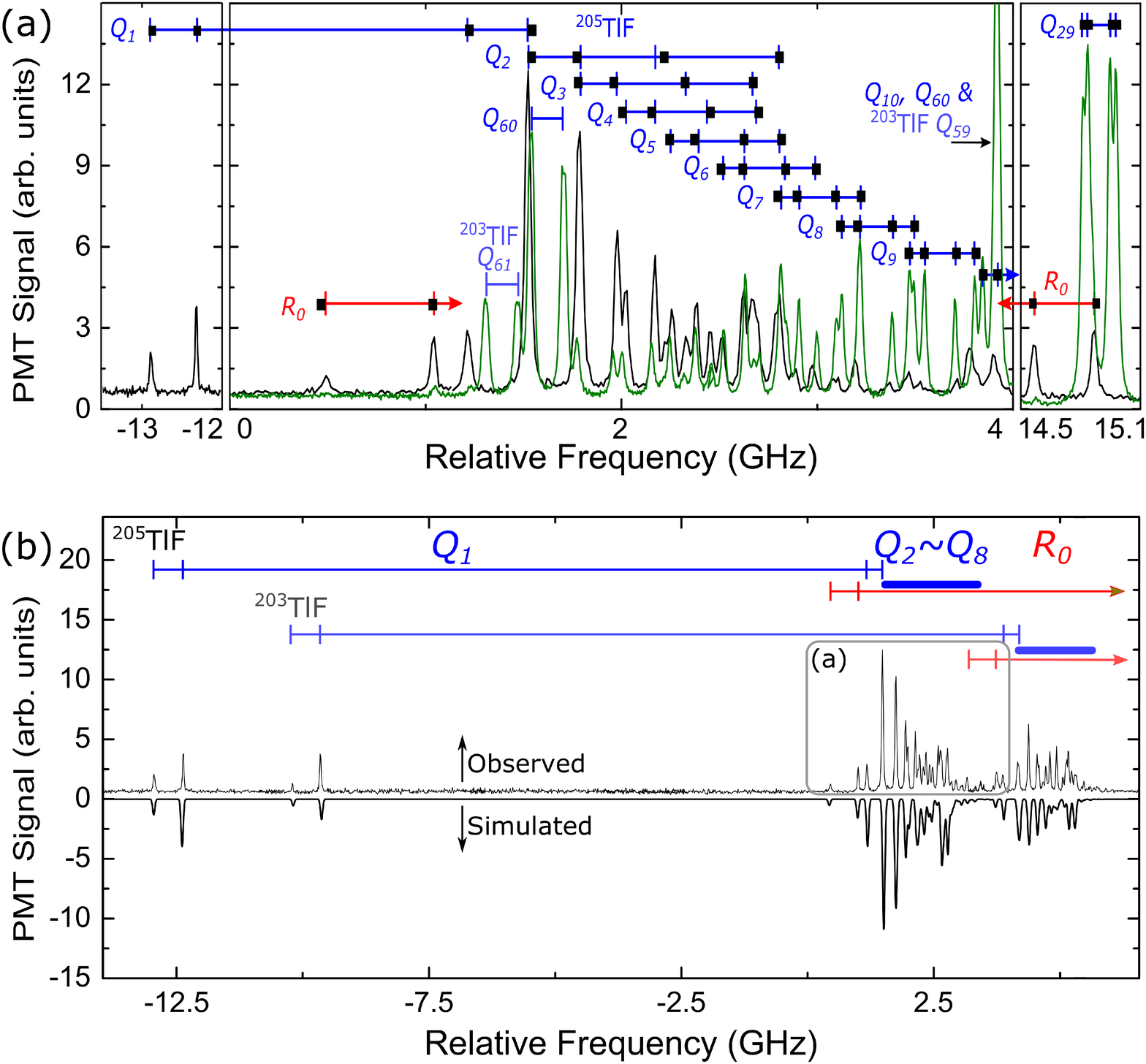}
        \caption{(Color online)  PMT signal vs laser frequency. (a) PMT signals from the thermal (green) and cryogenic (black) beams.  Comparison of the two spectra allow for tentative $^{205}$TlF line assignments (blue ticked lines) despite overlap of several low $Q$ branch lines  and $R_0$ (red ticked lines). These assignments are in excellent agreement with the line centers predicted by the best-fit Hamiltonian (black squares), even out to large values of $\Tilde{J}$, such as $Q_{29}$. The small variations of the line locations observed in the two sources is likely because the frequency data in the cryogenic source was not corrected for nonlinearities in the Fabry-Perot cavity scan. (b)  A scan over the $Q$ branches and part of $R_0$ in the cryogenic source.  Grey box approximates the frequency range of the central frame of panel (a).  While the HF structure of $\Tilde{J}=1$  covers $\sim 13$\,GHz, all higher $Q$ branch lines observable in the cryogenic source ($Q_2$ to roughly $Q_8$) are contained in only $\sim$\,2\,GHz (solid blue bar).  The large splitting in $\Tilde{J}=1$ means that for each Tl isotope, the lower two of the four $R_0$ lines are actually lower in frequency than the upper two $Q_1$ lines.  Below the observed spectrum is plotted an inverted simulated spectrum calculated from the fit Hamiltonian values in Table \ref{tab:hfstable}.  The full $1/e^2$ linewidth is set to the observed Doppler limited width of 31\,MHz. The simulated intensities are determined up to an overall scaling factor by the calculated relative line strengths $S$ \cite{Townes1955}, isotopic abundance, and an assumed Boltzmann rotational distribution with temperature $T$.   A fit to 10 lines representing $J_g$\,$=$\,$0$--4 gives $T = 3.6^{+1.0}_{-0.8}$\,K.  In the simulated spectrum, $^{203}$TlF $Q$ branch lines are produced using a fitted isotope shift and the $^{205}$TlF $f$-parity HF parameters, with $h_1$($^{203}$Tl) $= h_1$($^{205}$Tl)$g_{203}/g_{205}$  .
}\label{fig:Qscan}
\end{figure*}

The HF splitting is larger than the separation of $Q$ branch transitions for $Q_1$--$Q_{10}$, and clear identification of the lines again becomes problematic.  Ignoring the splittings which are obscured by overlapping $Q$ branch lines eliminates nearly all $Q_1$--$Q_{10}$ data (except $Q_4$ and $Q_9$). However, superposing the data taken in this spectral region from both the thermal beam and the cryogenic beam provides additional information and leads to the tentative assignments shown in Fig. \ref{fig:Qscan}a.
  The resulting splittings for $f$-parity excited states are listed in Table \ref{tab:fparity}.

 %Intro/Background
    We now discuss a number of unusual features in the $B$ state HF structure.  Many of these observations are attributed to an unusually large value of $h_1$(Tl) (found to be $\approx$\,29\,GHz), which we believe may be the largest observed HF interaction in any diatomic molecule.

    %While we perform calculations in the Hund's case (c) basis out of convenience, the lowest rotational levels of the $B$ state are better described by the uncommon Hund's case ($\rm{a_\alpha}$) \cite{Townes1955}.
\subsection{Large Hyperfine Splitting for $\Tilde{J} =1$}\label{sec:largeJ1}
The large value of $h_1$(Tl) is best illustrated by the enormous $b$ splitting in $\Tilde{J} =1$ (Fig.\,\ref{fig:Qscan}b).   While the HF structure of $\Tilde{J}=1$ covers $\sim\!13$\,GHz, all higher $Q$ branch lines to $\Tilde{J} \ge 2$ observable in the cryogenic source ($Q_2$ to roughly $Q_8$) are contained in only roughly 2\,GHz.  The large splitting in $\tilde{J}=1$ means two of the four $R_0$ lines are actually lower in frequency than $Q_2$.

 %Observed Splittings
 \subsection{J-mixing and ``Extra'' Lines}\label{sec:jmixing}

 \begin{figure*}
\centering
\includegraphics[width= \textwidth]{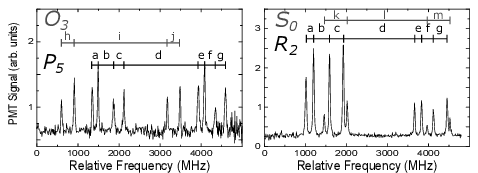}
        \caption{PMT signal vs laser frequency in the cryogenic source.  Examples of additional lines assigned to $O$ branch (left) and $S$ branch (right) transitions. }\label{fig:SObranch}
\end{figure*}

In addition to the expected $P$, $Q$ and $R$ branch lines, we observe additional quartets of lines near several low $P$ and $R$ branch lines (Fig.\,\ref{fig:SObranch}).  These are assigned as nominal $\Delta J$\,$=$\,$-2, +2$ transitions ($O$ and $S$ branch, respectively).   The presence of these lines is an indication that the $B$ state rotational levels are strongly mixed by the magnetic HF interaction.   If $\vert h_1$(Tl)$\vert$\,$\gg \vert h_1$(F)$\vert$, as expected since HF structure is enhanced in heavier species, then this $J$-mixing is only significant between states with the same values of $F_1$ \emph{and} $F$. The eigenstates can be written in the form
\begin{equation}
\begin{aligned}
  &\ket{\tilde{J},F_1=\tilde{J}+\frac{1}{2}} = \alpha_{F_1}\ket{J=\tilde{J}} +\beta_{F_1} \ket{J= \tilde{J}+1}, \\
   &\ket{\tilde{J}+1,F_1=\tilde{J}+\frac{1}{2}} = \beta_{F_1}\ket{J=\tilde{J}} -\alpha_{F_1} \ket{J= \tilde{J}+1}.
\end{aligned}
\end{equation}
Note that $\ket{\tilde{J}=1, F_1 = 1/2, F=0,1}$ is a special case which does not experience $J$-mixing.  For all other excited states, $J$-mixing leads to electric dipole-allowed $O$ and $S$ branch transitions.  Because the rotational constants of the $X$ and $B$ states are nearly identical, the $O_J$ lines appear very close to $P_{2J-1}$, and the $S_J$ lines appear very close to $R_{2J+2}$.  We label the splittings as $h,i,j$ for the $O$-branch and $k,l,m$ for the $S$-branch (see Fig.\,\ref{fig:SObranch}).

%Checks that Assignments make sense
 The $P$ and $R$ branches target the excited state $e$-parity levels.  We check our line assignments by comparing splittings in $R_J$ with those in $P_{J+2}$, which share a common excited state and thus should have the same observed HF splittings.   The $O$, $Q$, and $S$ branches target the excited state $f$-parity levels. The $Q_J$ branch HF structure must be compared to both $O_{J+2}$ and $S_{J-2}$ branches, as these lines address only one value of $F_1$ in the excited state due to the selection rule $\Delta F_1$\,$=$\,$\pm 1, 0$.

    %N,T branch?
     As an aside, we note that the $^{19}$F magnetic HF interaction mixes states with the same $F_1$, and thus it is in principle possible to drive electric dipole transitions with $\Delta J = -3, +3$, which we call the $N$ and $T$ branches, respectively.  For nearly equal ground and excited state rotational constants, we expect two  $T_J$ lines -- one for each isotope, and split by $d+e+f+g$ for $R_{J+2}$ -- to appear very close to $R_{3J+5}$ (and similarly, $N_J$ lines near $P_{3J-3}$).  Because the majority of the molecules produced in our cryogenic buffer gas beam source are in $J_g= 0,1$, we searched for $T_0$ and $T_1$ lines around $R_5$ and $R_8$, respectively.  However, we failed to detect any such transitions.

\subsection{Inverted $F_1$ Doublet for $\tilde{J} \ge 2$}

By examining the patterns in the data we can infer the ordering of the energy levels within different quartets of lines.  The $k$ splitting of the $S$ branch lines is consistent with the $c$  splitting (upper fluorine doublet splitting) of the $Q$ branch, indicating that $\ket{\tilde{J},F_1=\tilde{J} - \frac{1}{2}}$ lies higher in energy than $\ket{\tilde{J},F_1=\tilde{J} + \frac{1}{2}}$ for all observed $S$ branch lines, corresponding to $\tilde{J}$\,$=$\,$2$--5.  However, the $h$  splitting for $O_3$ is also consistent with the $c$ splitting of $Q_1$.  Combined with the fact that $a > c$ for all lines connected to $\tilde{J} =1$, but $a < c$ for $\Tilde{J} \ge 2$, we conclude that the ordering of $F_1$ levels is regular only in $\tilde{J}$\,$=$\,$1$, and inverted for $\tilde{J}$\,$\ge$\,2.  Also of note is that the $b$ splitting is $\sim\!29\times$ larger in $\tilde{J}$\,$=$\,$1$ than in $\tilde{J}$\,$=$\,$2$.  Naively, the $J$-scaling of diagonal matrix elements of $H^{\rm{eff}}_{\rm{mhf}}$ (Eq.\,\ref{eq:mhfI1}) would have led us to expect the $b$ splitting to only be $\sim 2\times$ larger in $\tilde{J}$\,$=$\,$1$ than in $\tilde{J}$\,$=$\,$2$.

  \begin{figure}[t]
\centering
\includegraphics[width= \linewidth]{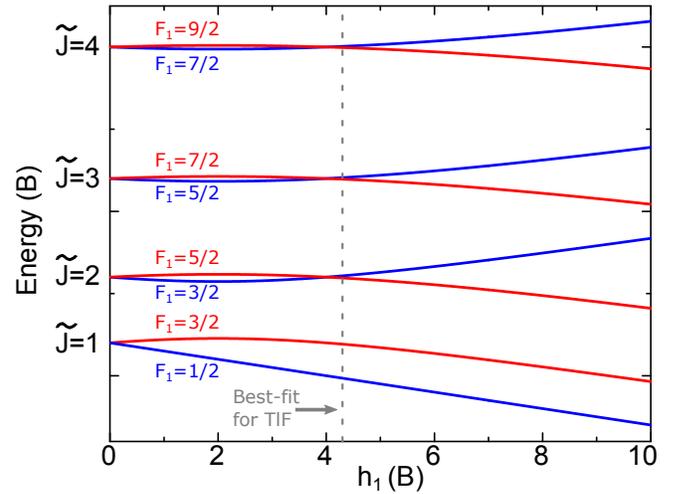}
        \caption{(Color Online) Eigenvalues of model Hamiltonian containing only rotation and magnetic HF interaction of a single nuclear spin $I=1/2$.  }\label{fig:orderswitch}
\end{figure}

\begin{table*}
  \centering
  \begin{ruledtabular}
  \begin{tabularx}{\linewidth}{l r r r|| r r r r }
  & \multicolumn{3}{c||}{This Work} & \multicolumn{4}{c}{Ref.\,\cite{Tiemann1988}}\\
  & \multicolumn{2}{c}{$e$-parity} & $f$-parity & \multicolumn{2}{c}{$e$-parity} &\multicolumn{2}{c}{$f$-parity} \\
 Parameter & \multicolumn{1}{c}{$^{203}$TlF}& \multicolumn{1}{c}{$^{205}$TlF}& \multicolumn{1}{c||}{$^{205}$TlF} & \multicolumn{1}{c}{$^{203}$TlF}& \multicolumn{1}{c}{$^{205}$TlF}& \multicolumn{1}{c}{$^{203}$TlF}& \multicolumn{1}{c}{$^{205}$TlF} \\ \hline
%& $^{203}$TlF, $e$-parity& $^{205}$TlF, $e$-parity& $^{205}$TlF, $f$-parity & $^{203}$TlF, $e$-parity& $^{205}$TlF, $e$-parity& $^{203}$TlF, $f$-parity& $^{205}$TlF, $f$-parity \\ \hline
$B_0$ & 6694.42(19) & 6689.09(16) & 6686.667(26) & 6694.5(11) & 6688.9(11) & 6694.3(11) &6688.7(11)\\
$D_0$ & & &0.010869(27) & 0.01089(14) &0.01088(14) &0.01098(14) & 0.01096(14)\\
$H_0\cdot10^8$ & & &-8.1(6) & -9.0(6) & -9.0(6)& -9.0(6)& -9.0(6)\\
& & & & & & \\
$h_1$(Tl) & 28516(5) &28793(5)& 28802(4)  \\
$h_1$(F) &864(1) & 859(1) & 871(1) \\
$c_I$(Tl) &-13.9(1) &-13.5(1) & -2.2(1) \\

\end{tabularx}
  \caption{Fit rotational and HF parameters for the $B^3\Pi_1$ state of TlF. Numbers in parentheses are the 1$\sigma$ confidence intervals.  All values are in MHz.
The root mean square deviation between our fitted and measured splittings is $<8$\,MHz in all cases.  The residual is $<20$\,MHz for each fitted point, and tends to decrease with increasing $\Tilde{J}$.}\label{tab:hfstable}
  \end{ruledtabular}
\end{table*}

  These peculiar features can be explained by an exceptionally large value of $h_1$(Tl). Consider a simplified system where only $H_{\rm{mhf}}$(Tl) (Eq.\,\ref{eq:mhf2}) and $H_{\rm{rot}}$\,$=$\,$B\boldsymbol{J}^2$ are present.  In the basis of states  $\ket{J=F_1\pm1/2, F_1}$ with $F_1 \ge 3/2$, the interaction is characterized by the $2\times2$ Hamiltonian
 \begin{equation}\label{eq:hamiltonian}
   H_{2\times 2} = \mqty(\frac{4F_1^2-1}{4}B+h &\frac{\sqrt{4F_1^2+4F_1-3}}{2(2F_1+1)}h \\
 \frac{\sqrt{4F_1^2+4F_1-3}}{2(2F_1+1)}h &\frac{(2F_1+3)(2F_1+1)}{4}B-h).
 \end{equation}
 By diagonalizing $H_{2\times 2}$, we find the energy $E_{F_1}^\pm$ of eigenstate $\ket{\tilde{J}=F_1\pm1/2, F_1}$ is given by
 \begin{equation}\label{eq:2x2energies}
   E_{F_1}^\pm = B(F_1+\frac{1}{2})^2\pm\frac{1}{2}\sqrt{B^2(2F_1+1)^2-4Bh+h^2}.
 \end{equation}
 The energy difference between states $\ket{\tilde{J},F_1=\tilde{J}\pm1/2}$ vanishes for  $\tilde{J} \ge 2$ when $h$\,$=$\,$4$$B$ (see Fig.\,\ref{fig:orderswitch}).  The energy ordering of the $F_1$ doublets reverses for $h_1$\,$>$\,$4 B$ for all $\tilde{J}\ge2$, but not for $\tilde{J}=1$.  For $h_1 \simeq 4 B$, the splitting of $\tilde{J}=1$ is much greater than that of all other states.  These features are consistent with our observations and our fit value $h_1\rm{(Tl)}$$/B_0$\,$=$\,$4.3 $.

\subsection{Fitting}

The best-fit values for the rotational and HF parameters are provided in Table \ref{tab:hfstable}.  The $B$ state rotational parameters are obtained by subtracting off the precisely-known $X$ state rotational energy from the weighted center of the HF quartet for each isotope, then fitting to a polynomial in $\Tilde{J}(\Tilde{J}+1)$. For the $f$-parity, we perform a cubic fit to $Q_{11}$--$Q_{60}$.  For the $e$-parity, we fit to $R_1$--$R_8$, and only the linear ($B_0$) term is statistically significant.  The $R_0$ lines are excluded from the fit as they are observed to deviate strongly from the rotational progression as discussed above. The $e$-parity level $B_0$ constants are observed to match the expected scaling with molecular reduced mass $\mu$: $^{203}B_0/^{205}B_0$\,$=$\,$1.00080(4)$, while $\mu_{205}/\mu_{203}=1.00084$.

The HF fit parameters are obtained by diagonalizing the effective Hamiltonian \makebox[\linewidth]{$H$\,$=$\,$H_{\rm{rot}}+$$H_{\rm{mhf}}(\rm{Tl})+$$ H_{\rm{mhf}}(\rm{F})+$$H_{\rm{nsr}}(\rm{Tl})+$$H_{\rm{nsr}}(\rm{F})$ \hfill for} each value of $F$. The HF parameters for the $f$- ($e$)-parity states were fit to the observed $Q$ branch (average of the $P$ and $R$ branch) splitting, weighted by their assigned uncertainties.  In all cases, we find $c_I$(F) to be consistent with zero within an uncertainty of 0.1\,MHz.

The uncertainty in the splittings measured in the thermal beam is 3\,MHz. For the cryogenic beam data, the Fabry-Perot ramp was less linear, and we assign an uncertainty of 8\,MHz to these splittings. We fit to the thermal beam data if available, and the cryogenic beam data when not.  In the instances where lines were measured in both setups, the agreement is typically within 10\,MHz.
As a check of our $Q$ branch splitting assignments where one or both lines are degenerate with other $Q$ branch lines (Table \ref{tab:fparity} bold data), a fit to the $f$-parity levels was performed excluding these data. However, this did not change the fit parameters within the assigned uncertainty.  The excellent agreement between the data and the line centers predicted by the best-fit Hamiltonian (Fig.\,\ref{fig:Qscan}a) provides additional support for our tentative $Q$ branch line assignments.

There is fair agreement between the $^{205}$TlF and $^{203}$TlF isotopologues on the value of the $^{19}$F doublet splittings. The magnetic HF parameter $h_1$(Tl) should be proportional to the nuclear $g$-factor $g_N$, and we find excellent agreement for $e$-parity: $h_1(^{203}$Tl)/$h_1(^{205}$Tl)$\,$=$\,$0.9904(2), while $g_{203}/g_{205}$\,$=$\,$0.990260$ \cite{Lide1997}.  Hyperfine anomalies are known to occur at the $10^{-4}$ level in neutral Tl atoms \cite{Pendrill1995, Richardson2000}, but is outside the precision of this study.  %The measured hyperfine anomaly $\Delta \equiv [(h_1(^{205}$Tl)$g_{203})/(h_1(^{203}$Tl)$*g_{205}-1] = -5.4\times 10^{-4}$ is in good agreement with the value $-4.7(1.5)\times 10^{-4}$ measured in atomic Tl \cite{Richardson2000}.

Finally, we extract the electronic isotope shift $\Delta \nu_{\rm{el}}(203-205)$ for the $X$\,$\rightarrow$\,$B$ transition \cite{Knockel1984}.  By averaging the shift for each HF transition for all $P$ and $R$ branch lines and subtracting off the calculated rovibrational contributions, we find $\Delta \nu_{\rm{el}}(203-205)$\,$=$\,$+3309(9)$\,MHz.  Because we only measure two Tl isotopes, we are unable to deconvolve the mass- and field-shift contributions to this quantity \cite{Knockel1984}.

\section{Optical Cycling}
\subsection{Applications and Requirements}\label{sec:cyclingapplications}

\begin{table*}
\begin{center}
\begin{ruledtabular}
\begin{tabular}{c r r r r || c c c}

Nominal State Label& \multicolumn{4}{c||}{Calculated Admixture $\ket{J,F_1,F}$} & $r_{\tilde{J}\tilde{J}-2}$& $r_{\tilde{J}\tilde{J}}$ & $r_{\tilde{J}\tilde{J}+2}$\\ \hline
$\ket{\tilde{J}=1,F_1=\frac{1}{2},F=0}$ & $\ket{1,\frac{1}{2},0}$ & & & & & 1 &\\
& & & & & & &\\

$\ket{\tilde{J}=1,F_1=\frac{1}{2},F=1}$ &$+0.9996\ket{1,\frac{1}{2},1}$&$+0.0203\ket{1,\frac{3}{2},1}$&$-0.0180\ket{2,\frac{3}{2},1}$& & & 0.9998 &0.0002\\
$\ket{\tilde{J}=1,F_1=\frac{3}{2},F=1}$ &$+0.0267\ket{1,\frac{1}{2},1}$&$-0.8518\ket{1,\frac{3}{2},1}$&$+0.5232\ket{2,\frac{3}{2},1}$ & & & 0.836 &0.164\\
$\ket{\tilde{J}=2,F_1=\frac{3}{2},F=1}$ &$+0.0048\ket{1,\frac{1}{2},1}$&$+0.5235\ket{1,\frac{3}{2},1}$&$+0.8520\ket{2,\frac{3}{2},1}$& & 0.091 & 0.909 &\\
& & & & & & &\\

$\ket{\tilde{J}=1,F_1=\frac{3}{2},F=2}$ &$+0.8482\ket{1,\frac{3}{2},2}$&$-0.5294\ket{2,\frac{3}{2},2}$&$-0.0138\ket{2,\frac{5}{2},2}$&$+0.0064\ket{3,\frac{5}{2},2}$ & & 0.832 &0.168\\
$\ket{\tilde{J}=2,F_1=\frac{5}{2},F=2}$ &$-0.0104\ket{1,\frac{3}{2},2}$&$+0.0120\ket{2,\frac{3}{2},2}$&$-0.9353\ket{2,\frac{5}{2},2}$&$+0.3535\ket{3,\frac{5}{2},2}$ &4$\cdot10^{-5}$ & 0.929 &0.071\\
$\ket{\tilde{J}=2,F_1=\frac{3}{2},F=2}$ &$+0.5295\ket{1,\frac{3}{2},2}$&$+0.8482\ket{2,\frac{3}{2},2}$&$+0.0011\ket{2,\frac{5}{2},2}$&$-0.0103\ket{3,\frac{5}{2},2}$ &0.094& 0.906 &6$\cdot10^{-5}$\\
$\ket{\tilde{J}=3,F_1=\frac{5}{2},F=2}$ &$+0.0040\ket{1,\frac{3}{2},2}$&$+0.0085\ket{2,\frac{3}{2},2}$&$+0.3536\ket{2,\frac{5}{2},2}$&$+0.9353\ket{3,\frac{5}{2},2}$ &0.050& 0.950\\
& & & & & & &\\

$\ket{\tilde{J}=2,F_1=\frac{5}{2},F=3}$ &$+0.9341\ket{2,\frac{5}{2},3}$&$-0.3568\ket{3,\frac{5}{2},3}$&$-0.0100\ket{3,\frac{7}{2},3}$&$+0.0032\ket{4,\frac{7}{2},3}$ & &0.927 &0.073\\
$\ket{\tilde{J}=3,F_1=\frac{7}{2},F=3}$ &$-0.0084\ket{2,\frac{5}{2},3}$&$+0.0074\ket{3,\frac{5}{2},3}$&$-0.9638\ket{3,\frac{7}{2},3}$&$+0.2665\ket{4,\frac{7}{2},3}$&3$\cdot10^{-5}$ & 0.961 &0.039\\
$\ket{\tilde{J}=3,F_1=\frac{5}{2},F=3}$ &$+0.3568\ket{2,\frac{5}{2},3}$&$+0.9341\ket{3,\frac{5}{2},3}$&$+0.0017\ket{3,\frac{7}{2},3}$&$-0.0084\ket{4,\frac{7}{2},3}$ &0.051 & 0.949 &4$\cdot10^{-5}$\\
$\ket{\tilde{J}=4,F_1=\frac{7}{2},F=3}$ &$+0.0023\ket{2,\frac{5}{2},3}$&$+0.0073\ket{3,\frac{5}{2},3}$&$+0.2666\ket{3,\frac{7}{2},3}$&$+0.9638\ket{4,\frac{7}{2},3}$&0.031 & 0.970\\

\end{tabular}
\caption{Eigenstates and their rotational branching ratios $r_{\tilde{J} J_g}$ for $\ket{B(v_e=0), F=0-3, P=f}$, calculated from the best-fit spectroscopic constants in Tables \ref{tab:fparity} and \ref{tab:hfstable} for $^{205}$TlF.}
\label{tab:admixture}
\end{ruledtabular}
\end{center}
\end{table*}

The ability to scatter many photons per molecule can be used for efficient detection and for application of optical forces.  The number of optical cycles required  before leaking into an uncoupled state depends on the application.   For instance, simple state-selective detection of molecules in a beam  often  does not require a highly closed cycle.
A typical laser-induced fluorescence setup can achieve overall detection efficiency for emitted photons (including detector quantum efficiency) of $\eta \sim 1$--10\% \cite{Barry2014}.  Hence, to detect molecules with near-unit efficiency,
 we only require an optical cycling scheme closed to (very roughly) $1/\eta\!\sim\!10$--100 cycles.  Applying significant optical forces requires a more closed cycle: $\sim\! 10^3$  optical cycles are required to achieve transverse Doppler cooling of a molecular beam to milliKelvin temperatures \cite{Shuman2010}, and $\sim\! 10^4$ optical cycles are needed to slow a molecular beam to a near stop \cite{Barry2012} or to magneto-optically trap molecules \cite{Barry2014}. % From binomial statistics, if the probability of decaying to any state outside of the optical cycle is $P$, then $1/e$ of the molecules remain after $1/P$ cycles.

 In this section, we consider vibrational and rotational branching effects which can limit the number of photons scattered on a given spectral line of the TlF $X$\,$\rightarrow$\,$B$ transition. We denote the vibrational branching fraction $v_e$\,$\rightarrow$\,$v_g$ by $b_{v_e v_g}$, and rotational branching fraction $\tilde{J}$\,$\rightarrow$\,$J_g$ (irrespective of vibrational quantum number) by $r_{\tilde{J} J_g}$.  We then propose a few optical cycling schemes which can be used in TlF for specific applications.

\subsection{Rotational Branching}

As discussed in Section \ref{sec:jmixing}, HF interactions mix states of the same total angular momentum $F$.
 In Table \ref{tab:admixture}, we present the eigenstates of $B(v_e$\,=\,$0)$ in the basis of Eq.\,\ref{eq:vectors} calculated from the best-fit spectroscopic constants in Table \ref{tab:hfstable}.
With the state admixture in hand, it is straightforward to calculate rotational branching fractions $r_{\tilde{J} J_g}$ using the line strengths $S$ \cite{Townes1955} for each excited basis state.

\subsection{Vibrational Branching}\label{sec:brancingratios}
Previous measurements of vibrational branching in TlF found favorable branching fractions $b_{v_e v_g}$ for cycling from the $B(v_e=0$ state:  $b_{00}$\,=\,$0.99$, $b_{01}$\,$<$\,$0.0002$, and $b_{02}$\,=\,$0.0110(6)$ \cite{Hunter2012}. Measurements of branching fractions  to $v_g$\,$\ge$\,3 were limited by experimental sensitivity and the availability of narrow bandpass interference filters at the needed wavelengths. Predicted values for these branching fractions using Morse and Rydberg-Klein-Rees (RKR) potentials, given in Table~\ref{tab:branchingratios}, were uncertain at the level required to determine which of these transitions could be neglected during longitudinal cooling. Here we present precision measurements of $b_{03}$, $b_{04}$, $b_{05}$, and $b_{06}$.

\begin{table}[t]
\begin{center}
\begin{ruledtabular}
\begin{tabular}{c c c c}
$B(0)\rightarrow X(v_g)$ & Morse  & RKR  & Measured\\ \hline
$v_g$=0 (271.7 nm) & 0.9892(3) & 0.989(2) & 0.989(2)\\
$v_g$=1 (275.3 nm) & 0.0003(2) & 0.0005(3)&$15(4)\cdot10^{-5}$\\
$v_g$=2 (278.8 nm) & 0.0104(2) & 0.010(2)& $0.011(2)$\\
$v_g$=3 (282.5 nm) & 0.00000(1) & $<$0.0003& $3(2)\cdot10^{-5}$\\
$v_g$=4 (286.2 nm) & 0.00013(1) & $<$0.0002& $13(3)\cdot10^{-5}$\\
$v_g$=5 (290.0 nm) & 0.00000(1) & $<$0.0003  & $1(1)\cdot10^{-5}$\\
$v_g$=6 (293.8 nm) & $-$ & $<$0.0002& $1(2)\cdot10^{-5}$\\

\end{tabular}
\caption[Predicted branching fractions $b_{0v_g}$]{Predicted and measured values of the branching fractions $b_{0v_g}$ in TlF. The calculations were done separately using the Morse potential and the RKR potential as models for the internuclear potential.}
\label{tab:branchingratios}
\end{ruledtabular}
\end{center}
\end{table}

Measurements of the branching fractions are done using a modified version of the thermal beam in order to take advantage of the  better molecular beam intensity stability. Pulsed excitation is used, and fluorescence is only detected after the laser pulse, eliminating scattered-light backgrounds.  The exciting laser light is produced by a pulsed dye laser with Coumarin 540A dye pumped by a 355\,nm Nd:YAG laser with pulse duration of 10\,ns and 10\,Hz repetition rate. The output of the dye laser is frequency doubled to produce 271.7\,nm light.	Measurements are made with the broadband pulsed system tuned to the largest fluorescence signal that occurs, near a large pile-up of rotational transitions between $Q_{30}$ and $Q_{52}$ with a bandhead that occurs at about $Q_{43}$. Background measurements are made by tuning to the high frequency side of this bandhead where virtually no fluorescence occurs.

\begin{figure*}
\centering
\includegraphics[width= .9\textwidth]{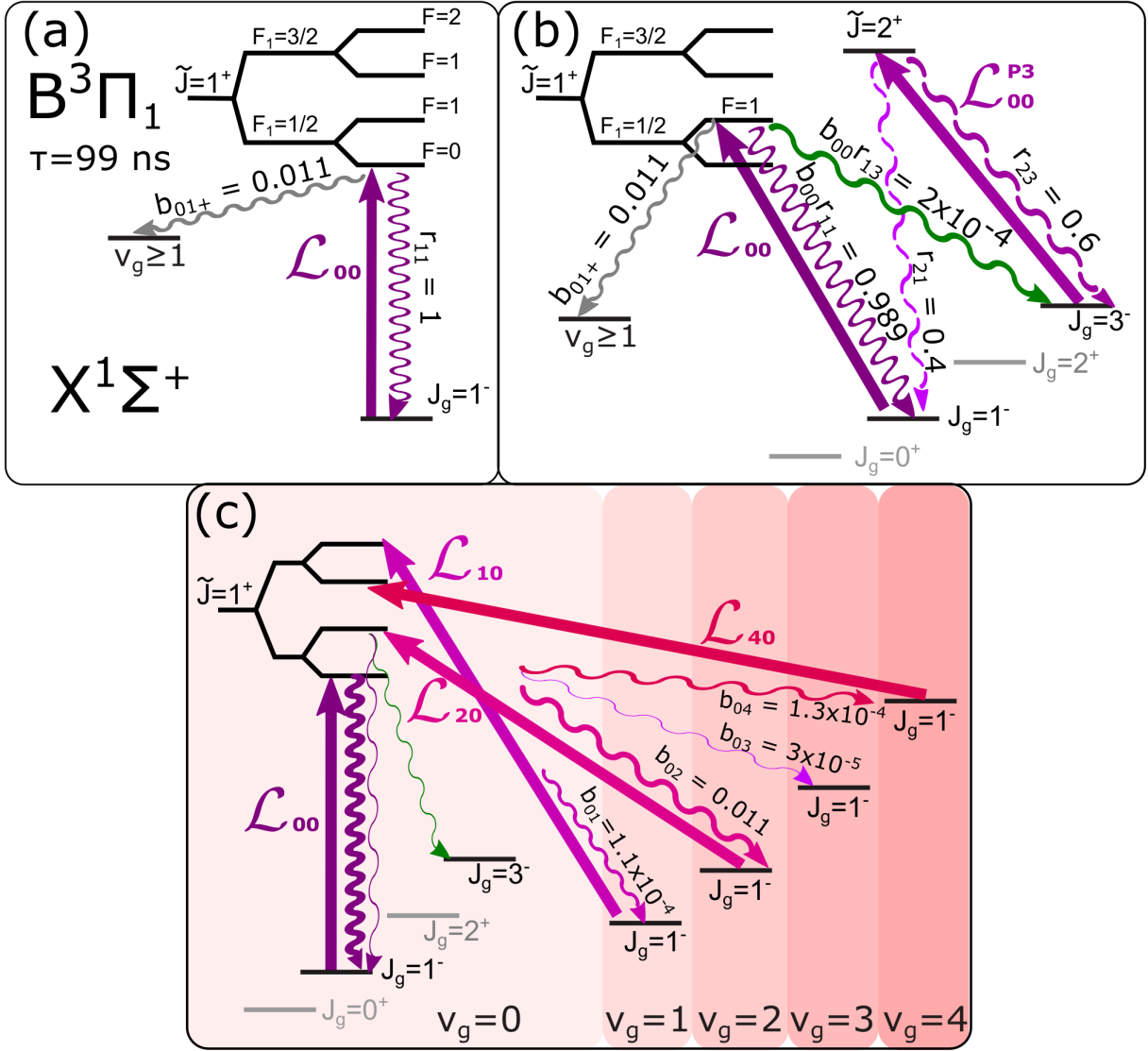}
        \caption{(Color Online) Optical cycling schemes for TlF. Straight arrows with labels $\mathcal{L}_{v_g v_e}$ denote laser excitations $X^1\Sigma^+(v_g)$\,$\rightarrow$\,$B^3\Pi_1(v_e)$, wavy arrows denote spontaneous decay.  Vibrational branching fractions are labeled $b_{v_e v_g}$, while rotational branching fractions for diagonal vibrational transitions are labeled as $r_{J^\prime J}$. (a) Cycling with $\ket{\Tilde{J}=1,F_1=1/2,F=1,P=+}$ as the excited state.  (b) Cycling with $\ket{\Tilde{J}=1,F_1=1/2,F=1,P=+}$ as the excited state.  Due to $J$-mixing, $\sim\!1/5000$ decays are to $J_g=3$.  This can be repumped to $J_g=1$ via a laser tuned to the $P_3$ transition, denoted $\mathcal{L}_{00}^{P3}$. (c) One of several possible optical cycling schemes closed at the level of $\sim\!10^4$ photon scatters.  The resolved HF structure of the excited state allows for repumping excited vibrational levels through $v_e\!=\!0$.  }\label{fig:cyclingdiagrams}
\end{figure*}

 Bandpass interference filters  centered at the wavelengths corresponding to the $B(0)$\,$\rightarrow$\,$X(0$--6$)$ transitions are placed before the ``signal'' PMT (above the vacuum chamber) to isolate fluorescence from the different vibrational bands.  A second ``normalization'' PMT (below the  chamber) contains only 271.7\,nm interference filters and is used to measure fluorescence from the main  transition ($B(0)$\,$\rightarrow$\,$X(0)$) at all times. This allows us to eliminate  fluctuations in laser and molecular beam intensity from our measurements.  To avoid saturation problems with the photon counting, we limit the average number of photons collected by either PMT to be $< 1$ per laser pulse.

  To measure a particular branching ratio we alternate between the corresponding filters, and count photons after subtracting off background fluorescence, normalizing with respect to the calibration signal, and taking into account the filter transmissions.
  Measurement of the ratio $b_{02}/b_{00}$  is accomplished with a modest molecular beam flux by  dividing the fluorescence signals when alternately filtering for $B(0)$\,$\rightarrow$\,$X(2)$ and $B(0)$\,$\rightarrow$\,$X(0)$.  However, the small rate of the decay to $v_g \ne 0,2$ requires higher molecular beam fluxes  to achieve adequate statistical precision.    To avoid saturating the normalization PMT from  these larger fluorescence signals,  additional attenuating filter  are added.   To avoid saturating the signal detector at these higher fluxes, all branching fractions $b_{0v}$ for $v_g \ne 0,2$ are measured by comparing the normalized signal for $B(0)$\,$\rightarrow$\,$X(v_g)$ to that of $B(0)$\,$\rightarrow$\,$X(2)$.  Their measured ratio is then multiplied by $b_{02}$ as measured above to determine $b_{0v_g}$.  The attenuation on the normalization transition is chosen so that the statistical significance of the measurement remains dominated by the number of photons counted in the signal detector.

We calculate the experimental branching fractions displayed in Table \ref{tab:branchingratios} using the assumption that the measured branching fractions account for all of the significant vibrational branches from $v_e$\,$=$\,$0$.  It is worth noting that the uncertainty associated with $b_{02}$ is larger than that quoted in our earlier publication \cite{Hunter2012}.  The increased uncertainty is due to the identification of a systematic error associated with a changing amount of light reflected into the normalization PMT when the interference filter in the signal PMT is switched between monitoring the $v_g=0$ and $v_g=2$ transitions.  This effect was likely also present in our earlier measurement.

\subsection{Optical Cycling Schemes}

We now examine a few specific, useful examples of  optical cycling $\ket{X^1\Sigma^+,J_g}$\,$\rightarrow$\,$\ket{B^3\Pi_1, \tilde{J}, F_1, F, P = f}$ (which we abbreviate below as $\ket{J_g}$\,$\rightarrow$\,$\ket{\tilde{J}, F_1, F}$).  First consider $\ket{ 1}$\,$\rightarrow$\,$\ket{1,1/2,0}$ (Fig.\,\ref{fig:cyclingdiagrams}a).  As there are no other $F$\,$=$\,$0$ states, the upper state of this transition is unmixed and all quantum numbers are exact to a very high extent (Table \ref{tab:admixture}).  Electric dipole and parity selection rules then dictate that the excited state can only decay to $J_g=1$.

Now consider $\ket{1}$\,$\rightarrow$\,$\ket{1,1/2,1}$ (Fig.\,\ref{fig:cyclingdiagrams}b).  This transition may be of interest in applications where a high photon scattering rate is desirable, as the higher excited state degeneracy $2F+1$ is expected to allow for roughly $3\times$ higher photon scattering rate than the case $\ket{ 1}$\,$\rightarrow$\,$\ket{1,1/2,0}$ \cite{Norrgard2015, NorrgardThesis}.  However, the $\ket{1,1/2,1}$ excited state has $\vert0.0180\vert^2 \simeq 3\times10^{-4}$ fractional $J=2$ character (Table \ref{tab:admixture}).  Decays from $J$\,$=$\,$2$ go to the desired $J_g=1$ (2/5 of the time) or to $J_g=3$ (3/5 of the time).  Hence, only $1/r_{13} \simeq 5000$ photons can be scattered before molecules are lost to the uncoupled $J_g=3$ state.  This loss is inconsequential for molecule detection applications, but is unacceptable for laser slowing and cooling.  This population could be recovered by repumping $J_g=3$ with an additional laser tuned to the $\ket{3}$\,$\rightarrow$\,$\ket{2,3/2,1,e}$ transition.  In this case, all electric dipole decay paths to $v_g$\,=\,0 would again be optically coupled.

Several possible optical cycling schemes exist which are closed to $\sim\!10^4$ photon scatters, sufficient for laser slowing or trapping.  We present one such scheme in Fig.\,\ref{fig:cyclingdiagrams}c.  The main cycling transition is $\ket{1}$\,$\rightarrow$\,$\ket{1,1/2,0}$ to minimize rotational branching.
Lasers $\mathcal{L}_{v_g v_e}$ (corresponding to the transition $X(v_g)$\,$\rightarrow$\,$B(v_e)$)  may be added to the optical cycle in order of decreasing importance ($\mathcal{L}_{00}$, $\mathcal{L}_{20}$, $\mathcal{L}_{40}$, $\mathcal{L}_{10}$), until the system is sufficiently closed for the intended application.  The resolved excited state HF structure allows the three strongest off-diagonal vibronic decays to be repumped through $v_e\!=\!0$ to take advantage of the near-unity $b_{00}$ branching fraction. We calculate that with only the $\mathcal{L}_{00}$ laser, $\sim$\,90 optical cycles may be achieved before molecules decay into a higher vibrational level.  This should be sufficient for high-efficiency detection. The addition of one repump laser $\mathcal{L}_{20}$ should be sufficient to achieve transverse cooling (closed to $\sim 3600$ cycles). All four lasers shown in Fig.\,\ref{fig:cyclingdiagrams}c would be necessary to achieve laser slowing or trapping.  In this scheme, we expect to scatter $\sim\! 10^4$ photons before decaying to unpumped levels.

\section{Conclusions}

In addition to spectroscopically identifying the $Q_1$ cycling transition of TlF, we have characterized the HF structure of the $B^3\Pi_1$ state.  Of particular note is the large magnetic HF interaction of the Tl nuclear spin.  With the exception of the lowest rotational and HF sublevel, the HF interactions significantly mix neighboring rotational levels of the $B$ state, and lead to additional rotational branching.

TlF appears to be an excellent molecule for optical cycling.  In particular, rotational and vibrational branching fractions presented here indicate  $\sim\!90$ optical cycles may be achieved using a single laser.  Efficient detection with a single laser should be possible in a symmetry violation measurement using TlF.  In addition, laser cooling and trapping should be feasible with four or fewer lasers.

While the measurements presented here show that the $X$\,$\rightarrow$\,$B$ system of TlF should allow for highly closed optical cycling, a number of considerations should be accounted for when choosing an appropriate cycling scheme in order to achieve a high photon scattering rate.  We plan to detail these considerations in a future paper.

\hfill\break
\indent Financial support was provided by the Army Research Office, the John Templeton Foundation, the Heising-Simons Foundation, and by NSF Grants No. PHY1205824 and PHY1519265.  The authors thank M. Kozlov and T. Steimle for helpful discussions.  S.S.A. and N.W. would like to thank Amherst College for Summer Research Awards.

\newpage

\section*{Appendices}
\subsection{Vibrational Branching Further Details}

In measuring the branching ratios in Section \ref{sec:brancingratios}, we use a number of interference filters to isolate fluorescence from individual vibronic bands.  Table \ref{tab:filter transmissions} provides further details on the filters used to monitor each transition.

\begin{table*}[t]
\centering
\begin{ruledtabular}
\begin{tabularx}{.6\linewidth}{c c c c c c}
$B(0)\rightarrow X(v_g)$ & Transition Wavelength $\lambda$ & Part Number &Transmission at $\lambda$ & Peak Wavelength & FWHM \\
\hline
 & & & & & \\
 $v_g=0$ & 271.7 nm & R214-01 &17.9\% & 271.8 nm & 10.3 nm\\
 & &R214-02 & 17.0\% & 271.3 nm & 10.1 nm\\
  & & neutral density&13.5\% & $-$ & $-$ \\
 & & & & & \\
 $v_g=1$ & 275.3 nm & T144-02 &14.1\% & 275.4 nm & 1.3 nm \\
 & & R088-01 &13.8\% & 275.3 nm & 1.3 nm\\
 & & & & & \\
 $v_g=2$ & 278.8 nm &S078-02 & 9.9\% & 279.3 nm & 2 nm\\
 & & & & & \\
 $v_g=3$ & 282.5 nm & T236-04 &12.9\% & 282.7 nm & 1.4 nm \\
 & & T236-05 & 12.4\% &282.8 nm & 1.5 nm \\
 & & & & & \\
 $v_g=4$ & 286.2 nm & T338-01 &11.4\% & 286.5 nm & 1.6 nm \\
 & & T335-10 &11\% & 290.9 nm & 10.7 nm \\
 & & & & & \\
 $v_g=5$ & 290.0 nm & T338-02 &11\% & 290.3 nm & 1.6 nm \\
 & & T335-10 &17.8\% & 290.9 nm & 10.7 nm\\
 & & & & & \\
 $v_g=6$ & 293.8 nm & T335-14 &11.7\% & 297.9 nm & 10.8 nm \\
& & T335-15 & 10.0\% & 298.7 nm & 11.6 nm \\
 & & & & & \\
\end{tabularx}
\caption{Part numbers (Andover), transmissions, and bandwidths of all the bandpass filters used in this experiment.  Uncertainty in the transmissions of all filters except the neutral density filter is 1.5\%. Uncertainty in transmission of the neutral density filter at the given wavelength is 1\%. Uncertainties were chosen to reflect the range of possible transmissions that could result with incidences between $\pm$ 2$^\circ$ off normal.}
\label{tab:filter transmissions}

\end{ruledtabular}

\end{table*}

\subsection{Racah Algebra Identities}

  Here we provide a number of useful identities and their equation numbers in Ref.\,\cite{Brown2003}.  In the following, $j_1$ and $j_2$ are general angular momenta which couple to form angular momentum $j_{12}$.  Rank-$k$ tensor operators $T^k(\boldsymbol{A_1})$ and $T^k(\boldsymbol{A_2})$ act on $j_1$ and $j_2$, respectively.  Tensors are represented in the spherical basis, with index $p$ ($q$) for the lab frame (molecule-fixed frame) coordinate system.

\emph{Transformation from molecule-fixed frame to lab frame:}
\begin{equation}\label{eq:BC5144}\tag{5.144}
\begin{aligned}
  T^k_q(\boldsymbol{A}) &= \sum_{p} \mathscr{D}^{(k)}_{pq}(\omega)T^k_p(\boldsymbol{A})\\
  &= \sum_{p} (-1)^{p-q} \mathscr{D}^{(k)}_{-p-q}(\omega)^*T^k_p(\boldsymbol{A}).
  \end{aligned}
\end{equation}

\emph{Scalar product of two tensor operators:}
\begin{equation}\label{eq:BC5173}\tag{5.173}
\begin{aligned}
  &\matrixelement{j_1,j_2,j_{12},m}{T^k(\boldsymbol{A_1})\cdot T^k(\boldsymbol{A_2})}{j_1^\prime,j_2^\prime,j_{12}^\prime,m^\prime} \\
   &\quad= (-1)^{j_1^\prime+j_{12}+j_2}\delta_{j_{12},j_{12}^\prime}\delta_{m,m^\prime} \begin{Bmatrix}
                                                                                 j_2^\prime & j_1^\prime & j_{12} \\
                                                                                 j_1 & j_2 & k
                                                                               \end{Bmatrix}\\
                                                                                &\quad\quad \times \matrixelement{j_1}{\vert T^k(\boldsymbol{A_1}) \vert}{j_1^\prime}\matrixelement{j_2}{\vert T^k(\boldsymbol{A_2}) \vert}{j_2^\prime}.
                                                                               \end{aligned}
\end{equation}

\emph{Spectator Theorem:}
\begin{equation}\label{eq:BC5174}\tag{5.174}
\begin{aligned}
  &\matrixelement{j_1,j_2,j_{12}}{\vert T^k(\boldsymbol{A_1})\vert}{j_1^\prime,j_2^\prime,j_{12}^\prime} \\ &\quad =(-1)^{j_{12}^\prime +j_1+k+j_2}\delta_{j_2,j_2^\prime}\sqrt{(2j_{12}+1)(2j_{12}^\prime+1)}\\
  &\quad\quad\times \begin{Bmatrix}
                                                   j_1^\prime & j_{12}^\prime & j_2 \\
                                                   j_{12} & j_1 & k
                                                 \end{Bmatrix} \matrixelement{j_1}{\vert T^k(\boldsymbol{A_1}) \vert}{j_1^\prime}.
\end{aligned}
\end{equation}

\emph{Reduced matrix element of a rank-1 tensor operator:}
\begin{equation}\label{eq:BC5179}\tag{5.179}
  \matrixelement{j_1}{\vert T^1(\boldsymbol{A_1}) \vert}{j_1^\prime} = \delta_{j_1,j_1^\prime}\sqrt{j_1(j_1+1)(2j_1+1)}.
\end{equation}

\emph{Reduced matrix element of the partially reduced Wigner rotation matrix:}
\begin{equation}\label{eq:BC5186}\tag{5.186}
\begin{aligned}
 &  \matrixelement{J,\Omega}{\vert \mathscr{D}^{(k)}_{\cdot,q}(\omega)^* \vert}{J^\prime,\Omega^\prime}\\
  &\quad = (-1)^{J-\Omega}\sqrt{(2J+1)(2J^\prime+1)}\mqty(J&k&J^\prime\\-\Omega&q&\Omega^\prime).
  \end{aligned}
\end{equation}

\subsection{Evaluating Matrix Elements}\label{sec:MatrixElements}
Here we derive the HF matrix elements given in Section \ref{sec:hyperinehamiltonian}.
We have found references \cite{Brown2003,Okabayashi2003,Okabayashi2012,Knurr2009} helpful in understanding which terms are likely important in the HF structure Hamiltonian.  Because Ref.\,\cite{Brown2003} contains all necessary fomulae for Racah algebra for diatomic molecules, we use the short-hand $\underset{(x.xx)}{=}$ for ``is equivalent by Ref.\,\cite{Brown2003} Eq.\,(x.xx)''.

   We begin by rotating the operator  of Equation \ref{eq:mhf2} into the lab frame:
 \begin{equation}\label{eq:mhf3}
 \begin{aligned}
  &\matrixelement{J,\Omega,F_1,F,m}{T_{q=0}^1(\boldsymbol{I})}{J^\prime,\Omega^\prime,F_1^\prime,F,m} \underset{(5.144)}{=}  \\
  &\sum_{p} \matrixelement{J,\Omega,F_1,F,m}{\mathscr{D}^{(1)}_{-p,0}(\omega)^* T_{p}^1(\boldsymbol{I})}{J^\prime,\Omega^\prime,F_1,F,m}
\end{aligned}
\end{equation}
For $I = I_1$,
\begin{equation}\label{eq:mhf4}
  \begin{aligned}
  &\underset{(5.173)}{=} (-1)^{J^\prime +F_1 +I_1}\delta_{F_1,F_1^\prime} \begin{Bmatrix}
I_1 & J^\prime & F_1 \\
J & I_1 & 1
\end{Bmatrix}\\
&\quad\quad\times
 \matrixelement{J,\Omega}{\vert \mathscr{D}^{(1)}_{\cdot,0}(\omega)^* \vert}{J^\prime,\Omega^\prime}
  \matrixelement{I_1}{\vert T^1(\boldsymbol{I_1}) \vert}{I_1}\\
  &\underset{\substack{(5.179)\\(5.186)}}{=} (-1)^{J+J^\prime +F_1 +I_1-\Omega}\delta_{F_1,F_1^\prime} \\
  &\quad\quad\times  \begin{Bmatrix}
I_1 & J^\prime & F_1 \\
J & I_1 & 1
\end{Bmatrix} \mqty( J & 1 & J^\prime \\ -\Omega & 0 & \Omega^\prime) \\
&\quad\quad\times [(2J+1)(2J^\prime+1)I_1(I_1+1)(2I_1+1)]^{1/2} ,
  \end{aligned}
\end{equation}
which matches Ref.\,\cite{Okabayashi2012} Eq.\,(6). Note that when employing (5.173) to get the first line of Eq.\,\ref{eq:mhf4}, both the Wigner rotation matrix $\mathscr{D}$ and $T^1(\boldsymbol{I_1})$ are being reduced in the lab frame (indexed by $p$). Specifically, the term $\matrixelement{J,\Omega}{\vert \mathscr{D}^{(1)}_{\cdot,0}(\omega)^* \vert}{J^\prime,\Omega^\prime}$ is the partially reduced Wigner rotation matrix (reduced in the lab frame, but not the molecule-fixed frame), whose value is given by Ref.\,\cite{Brown2003} Eq.\,\ref{eq:BC5186}. For $I = I_2$, Eq.\,\ref{eq:mhf3} yields
\begin{equation}\label{eq:mhf5}
  \begin{aligned}
  &\underset{(5.173)}{=} (-1)^{F_1^\prime +F +I_2}\begin{Bmatrix}
I_2 & F_1^\prime & F \\
F_1 & I_2 & 1
\end{Bmatrix}\\
  &\quad\times
 \matrixelement{J,\Omega,F_1}{\vert \mathscr{D}^{(1)}_{\cdot,0}(\omega)^* \vert}{J^\prime,\Omega^\prime,F_1^\prime}
  \matrixelement{I_2}{\vert T^1(\boldsymbol{I_2}) \vert}{I_2}\\
   &\underset{(5.174)}{=} (-1)^{F_1^\prime +F +I_2}  \begin{Bmatrix}
I_2 & F_1^\prime & F \\
F_1 & I_2 & 1
\end{Bmatrix}\\
&\quad\times(-1)^{F_1^\prime+J+1+I_1}[(2F_1+1)(2F_1^\prime+1)]^{1/2} \begin{Bmatrix}
J^\prime & F_1^\prime & I_1 \\
F_1 & J & 1
\end{Bmatrix}\\
&\quad \times \matrixelement{J,\Omega}{\vert \mathscr{D}^{(1)}_{\cdot,0}(\omega)^* \vert}{J^\prime,\Omega^\prime}
  \matrixelement{I_2}{\vert T^1(\boldsymbol{I_2}) \vert}{I_2}\\
  &\underset{\substack{(5.179)\\(5.186)}}{=} (-1)^{2F_1^\prime+F+2J+1+I_1+I_2-\Omega}  \\
  &\quad \times \begin{Bmatrix}
I_2 & F_1^\prime & F \\
F_1 & I_2 & 1
\end{Bmatrix}\begin{Bmatrix}
J^\prime & F_1^\prime & I_1 \\
F_1 & J & 1
\end{Bmatrix}  \mqty( J & 1 & J^\prime \\ -\Omega & 0 & \Omega^\prime)\\
&\quad  \times \bigl[(2F_1+1)(2F_1^\prime+1)(2J+1)(2J^\prime+1)\\
&\quad\quad \times I_2(I_2+1)(2I_2+1)\bigr]^{1/2} ,
  \end{aligned}
\end{equation}
which again matches Ref.\,\cite{Okabayashi2012} Eq.\,(6).

\pagebreak

The nuclear spin-rotation matrix elements can be written as follows.  For $I = I_1$:\samepage{
 \begin{equation}\label{eq:nsr1a}
 \begin{aligned}
  &\matrixelement{J,\Omega,F_1,F,m}{T^1(\boldsymbol{I_1})\cdot T^1(\boldsymbol{J})}{J^\prime,\Omega^\prime,F_1^\prime,F,m} \\
  &\underset{(5.173)}{=} (-1)^{J^\prime+F_1+I_1}\delta_{F_1,F_1^\prime}\begin{Bmatrix}
I_1 & J^\prime & F_1 \\
J & I_1 & 1
\end{Bmatrix}\\
&\quad\quad\times \matrixelement{J}{\vert T^1(\boldsymbol{J})\vert}{J^\prime}
  \matrixelement{I_1}{\vert  T^1(\boldsymbol{I_1}) \vert}{I_1}\\
 &\underset{(5.179)}{=} (-1)^{J+F_1+I_1}\delta_{F_1,F_1^\prime}\delta_{J,J^\prime}\begin{Bmatrix}
I_1 & J & F_1 \\
J & I_1 & 1
\end{Bmatrix}\\
&\quad\quad\times[(J(J+1)(2J+1)I_1(I_1+1)(2I_1)+1)]^{1/2}.
\end{aligned}
\end{equation}
Now for $I = I_2$:
\begin{equation}\label{eq:nsr2a}
 \begin{aligned}
  &\matrixelement{J,\Omega,F_1,F,m}{T^1(\boldsymbol{I_2})\cdot T^1(\boldsymbol{J})}{J^\prime,\Omega^\prime,F_1^\prime,F,m} \\
  &\underset{(5.173)}{=} (-1)^{F_1^\prime+F+I_2}\begin{Bmatrix}
I_2 & F_1^\prime & F \\
F_1 & I_2 & 1
\end{Bmatrix}\\
&\quad\times \matrixelement{J,I_1,F_1}{\vert T^1(\boldsymbol{J})\vert}{J^\prime,I_1,F_1^\prime}
  \matrixelement{I_2}{\vert  T^1(\boldsymbol{I_2}) \vert}{I_2}\\
& \underset{(5.174)}{=} (-1)^{F_1^\prime+F+I_2}\begin{Bmatrix}
I_2 & F_1^\prime & F \\
F_1 & I_2 & 1
\end{Bmatrix}\\
&\quad\times(-1)^{F_1^\prime+J+I_1+1}[(2F_1+1)(2F_1^\prime)]^{1/2}\begin{Bmatrix}
J^\prime & F_1^\prime & I_1 \\
F_1 & J & 1
\end{Bmatrix}\\
&\quad\times \matrixelement{J}{\vert T^1(\boldsymbol{J})\vert}{J^\prime}
  \matrixelement{I_2}{\vert  T^1(\boldsymbol{I_2}) \vert}{I_2}\\
&\underset{(5.179)}{=} (-1)^{2F_1^\prime+F+J+I_1+I_2+1}\delta_{J,J^\prime}\\
&\quad\times \begin{Bmatrix}
I_2 & F_1^\prime & F \\
F_1 & I_2 & 1
\end{Bmatrix}\begin{Bmatrix}
J^\prime & F_1^\prime & I_1 \\
F_1 & J & 1
\end{Bmatrix}\\
&\quad\times \bigl[(2F_1+1)(2F_1^\prime+1)J(J+1)(2J+1)\\
&\quad\quad\times I_2(I_2+1)(2I_2+1)\bigr]^{1/2}.
\end{aligned}
\end{equation}}

Results \ref{eq:nsr1a} and \ref{eq:nsr2a} match Ref.\,\cite{Brown2003} (8.290) and (8.291), respectively, as well as Ref.\,\cite{Knurr2009} (5).  However, Eq.\,\ref{eq:nsr1a} differs from Ref.\,\cite{Okabayashi2012} by a factor of $[J(J+1)(2J+1)]^{1/2}$.

\subsection{Full $Q$ Branch Spectrum}\label{sec:Full Q branch spectrum}

\begin{sidewaysfigure}[p]
\includegraphics[width=\textwidth]{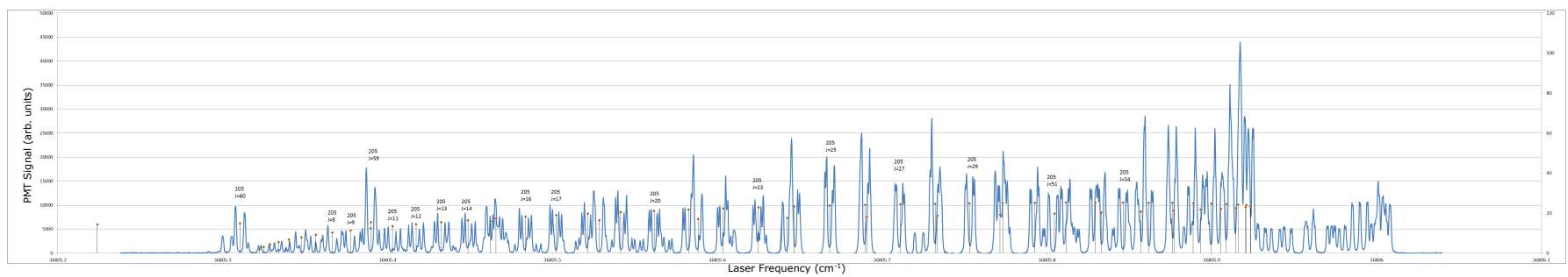}
\caption{PMT signal vs laser frequency in the thermal beam showing all identified $Q$ branch lines. Drop lines indicate calculated line centers from best-fit rotational constants (Table \ref{tab:hfstable}).}\label{fig:BigQScan}
\end{sidewaysfigure}

In Fig.\,\ref{fig:BigQScan}, we include the full $Q$ branch spectrum of the $X$\,$\rightarrow$\,$B$ transition in TlF.  Line centers predicted by the best-fit rotational constants in Table\,\ref{tab:hfstable} are in excellent agreement with the observed spectrum.

\hfill\break
\clearpage
% Finally, it is often helpful to know how matrix elements scale with rotational level, and the quantum numbers in which the Hamiltonian is diagonal.  These are provided for the effects considered in this paper in Table \ref{tab:Jtrends}.

%\begin{table}[b]
%\centering
%\begin{ruledtabular}
%\begin{tabularx}{.4\textwidth}{l c c c }
%\centering
% Parameter & Exact Diagonal Term & Leading Order $J$ Scaling &Diagonal Quantum Numbers\\ \hline
%$h_1$(Tl)& & $1/J$ & $F, F_1, \Omega, P$\\
%$h_1$(F) & & $1/J$ & $F, \Omega, P$\\
%$h_1q$(Tl) & & $1/J$ & $F, F_1, P$\\
%$h_{1D}$(Tl) & &$J$ & $F, F_1, \Omega, P$\\

%$c_I$(Tl) & &$J$ & $F, J, \Omega, P$\\
%$c_I$(F) &$c_I$(F)$J/2$ &$J$ & $F, F_1, \Omega, P$
%\end{tabularx}\caption{Scaling of the diagonal hyperfine terms with $J$.}\label{tab:Jtrends}
%\end{ruledtabular}
%\end{table}

\bibliography{thebib}

\clearpage

\end{document}